\documentclass[aps,prb,reprint,groupedaddress,showpacs]{revtex4-1}
\usepackage{amsmath}
\usepackage{amssymb}
\usepackage{graphicx}
\usepackage{epstopdf}
\usepackage{dcolumn}
\usepackage{bm}

\begin{document}

\title{Magnon Waves on Chains of YIG particles: Dispersion Relations, Faraday Rotation, and Power Transmission}

\author{Nicholas A. Pike}
\email{Nicholas.pike@ulg.ac.be}
\affiliation{Department of Physics, University of Liege, 4000 Liege, Belgium}
\affiliation{Department of Physics, The Ohio State University, Columbus, Ohio 43210, USA }%
\author{David Stroud}
\affiliation{ Department of Physics, The Ohio State University, Columbus, Ohio 43210, USA }%

\date{\today}

\begin{abstract}
We calculate the dispersion relations for magnon waves on a periodic chain of spherical or cylindrical Yttrium Iron Garnet (YIG) particles.  We use the quasistatic approximation, appropriate when $kd \ll 1$, where $k$ is the wave number and $d$ the interparticle spacing.   In this regime, because of the magnetic dipole-dipole interaction between the localized magnetic excitations on neighboring particles, dispersive magnon waves can propagate along the chain.   The waves are analogous to plasmonic waves generated by electric  dipole-dipole interactions between plasmons on neighboring metallic particles.   The magnon waves can be longitudinal ($L$), transverse ($T$), or elliptically polarized.   We find that a linearly polarized magnon wave undergoes a Faraday rotation as it propagates along the chain.  The amount of Faraday rotation can be tuned  by varying the off-diagonal component of the permeability tensor.   We also discuss the possibility of wireless power transmission along the chain using these coupled magnon waves.
\end{abstract}


\maketitle
\section{Introduction}

There have recently been many proposals for ways to induce energy propagation along a metallic nanoparticle chain.  Such propagation occurs in the form of traveling waves, which are driven by electric dipole-dipole coupling between localized plasmon modes on neighboring nanoparticles\cite{Maier2003,Brongersma2000}.   The dispersion relations of such waves have been calculated by several authors within the quasistatic approximation~\cite{Maier2003,Park2004,Pike2013,Pike2016}.  This approximation, which assumes that the dipole-dipole coupling is accurately described by electrostatics, is believed accurate when the spacing between particles is small compared to the wavelength of light at a typical plasmon frequency. The resulting propagating plasmonic waves have been observed in recent experiments~\cite{Hossain2012,Li2011}.  These calculations have also been extended to included dynamical effects beyond the quasistatic approximation~\cite{Weber2004}.

In the present paper, we investigate the analogous problem of waves propagating along chains of {\it magnetic} particles coupled via their magnetic dipole moments.   We consider particles of the well known magnetic insulator Yttrium Iron Garnet (YIG).   The magnetic dipole-dipole interaction between localized magnetic excitations on these particles leads to propagating magnetic waves, or magnons.  These waves can carry both magnetization and energy and thus may produce wireless power transmission~\cite{Karenowska2015,Damon1965,Kreisel2009}. Specifically, we calculate the dispersion relations of propagating magnon waves along chains of spherical or cylindrical YIG particles within the quasistatic approximation.  These calculations are analogous to the extensive earlier work, mentioned above, on wave propagation along chains of small metal particles, also within the quasistatic approximation.   In this latter work, the waves are able to propagate because of the electric dipole-dipole interactions between the plasmonic excitations in the individual nanoparticles.   By contrast, the relevant interactions for the YIG particles are {\it magnetic} dipole-dipole interactions. Here, we use the magnetic properties of YIG, together with equations of motion describing the interactions of localized magnetic excitations on the particles with the excitations on neighboring particles, to calculate the dispersion relations of propagating magnon waves along chains of spherical or cylindrical YIG particles.  

Recently, other types of magnon waves have been studied theoretically at the interface between a two-dimensional (2D) magneto-optical photonic crystal and a regular 2D gapped photonic crystal~\cite{Wang2008a}.  These studies show that one-way edge modes can propagate at the interface and within the band gap of the bulk modes.  They also show that these edge modes are confined to within a few lattice constants of the boundary between the two crystals, and are thus basically one-dimensional.   Work by Khanikaev {\it et al.}~\cite{Khanikaev2007} in metallic systems containing holes filled with the magneto-optical material Bi:YIG suggest that the Faraday rotation angle within the material is enhanced because of plasmonic excitations at the surface of the magneto-optic material.   

The goal of transmitting power wirelessly is, of course, very old. In the early $20^{th}$ century, Tesla~\cite{Tesla1914,Tesla1891} attempted to transmit power wirelessly even before the dawn of the electrical power grid.  His proposal, like many others~\cite{Li2011a,Javadi2013}, requires the generation of large electric fields in order for a reasonable amount of power to be transmitted.   In order to avoid the need for large electric fields, recent efforts have turned towards magnetic systems.  In particular, it has been shown~\cite{Kur2007} that one can use a coil of wire, in which a current is driven at a fixed frequency, to transmit power, via magnetic induction, to another coil of wire a few meters away, with approximately 40\% efficiency. Likewise, one can transmit power using split-ring resonators~\cite{Shiffler2013} or many other patterned devices. In the present work, we describe an approach which requires only single crystals of YIG~\cite{Kimura1977,Linares1965}, patterned as spheres or rods and arranged in a periodic chain, in order to produce wireless power transmission.

We turn now to the body of the paper.  In Section~\ref{model}, we describe the geometry of the system and derive equations for the coupled magnetic dipole moments within the quasistatic approximation, using an approach analogous to some earlier work~\cite{Brongersma2000,Maier2003,Pike2013}. In Section~\ref{three},  we write down the propagating wave solutions for chains of spherical or cylindrical YIG particles, and illustrate the solutions with some numerical results. Finally, in Section~\ref{four}, we discuss the results of our calculations and provide some concluding remarks.  A short Appendix provides further algebraic details about the calculation of the dispersion relations for chains of YIG spheres.

\section{Model}\label{model}

\begin{figure}[t]
\includegraphics[width=0.5\textwidth]{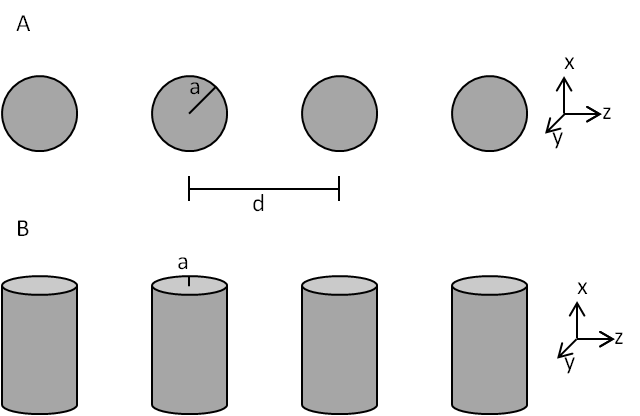} 
\caption{(a) Considered geometry for a chain of YIG spheres of radius $a$ and separation $d$ periodically arranged along the $z$ axis.  (b)  A chain of  very long YIG rods of radius $a$, arranged periodically along the $z$ axis with separation $d$. In both cases, we assume that the static magnetic field ${\bf H}_0$ and the saturation magnetization ${\bf M}_s$ are parallel to one another and lie along either the $x$ or $z$ axis.}
\label{fig:one}
\end{figure}

We will consider periodic chains of spherical or cylindrical YIG particles.  The spacing between the centers is denoted $d$, while the radius of the spheres or cylinders is $a$. For both spheres and cylinders, the chains run along the $z$ axis, while for the cylinders, the cylinder is parallel to the $x$ axis.  In all cases, we assume that $d = 3 a$. A cartoon image of the assumed geometries is shown in Fig.~\ref{fig:one}.

We will be interested in periodic waves of magnetic dipole moments traveling along the chain.  As shown elsewhere in the case of coupled electric dipole moments~\cite{Pike2013,Maier2001,Maier2003,Koenderink2006,Brongersma2000,Maier2003}, these magnetic dipole waves can be obtained from coupled linear equations involving the dipole moments on each particle within the quasistatic approximation.  For the electric dipole case, the dispersion relations are controlled by electric dipole-dipole coupling between plasmons centered on neighboring metallic particles.  In the case of YIG particles, the dispersion relations describe propagating waves of magnetic dipoles and are generated by magnetic dipole-dipole coupling between magnetic plasmon excitations centered on neighboring YIG particles.  The solutions to the coupled set of equations lead to the dispersion relations for the three branches of propagating magnon waves.

To obtain the coupled linear equations it is convenient to proceed using equations analogous to Eq.\ (9) of Ref.~\citenum{Pike2013}.   We assume that the chain of particles (either spheres or cylinders) is parallel to the $z$ axis as shown in Fig.~\ref{fig:one}. Writing out the coupled equations explicitly, and assuming, first, that the magnetic particles are spheres of radius $a$, we obtain
\begin{equation}\label{eq:mag_moment3D}
{\bf m}_n = -\frac{4\pi a^3}{3}\hat{t}\sum_{n^\prime \neq n}\hat{\cal G}^{s}({\bf x}_n - {\bf x}_{n^\prime})\cdot {\bf m}_{n^\prime},
\end{equation}
where ${\bf m}_n$ is the  magnetic dipole on  the n$^{th}$ sphere.  If the particles are cylinders of radius $a$, we find
\begin{equation}
\label{eq:mag_moment2D}
{\bf m}_n = -\pi a^2\hat{t}\sum_{n^\prime \neq n}\hat{\cal G}^{c}({\bf x}_n - {\bf x}_{n^\prime})\cdot {\bf m}_{n^\prime},
\end{equation}
where the quantity ${\bf m}_n$ is now the magnetic moment per unit cylinder length of the n$^{th}$ cylinder.

In Eqs.~\eqref{eq:mag_moment3D} and~\eqref{eq:mag_moment2D} we find that $\hat{\cal G}^{s}$ or $\hat{\cal G}^c$ are related to the Greens functions of a point charge in two or three dimensions.  Specifically, ${\cal G}_{ij}^s({\bf x}-{\bf x}^\prime) = \partial_i^\prime\partial_j G^s({\bf x}-{\bf x}^\prime)$, where $G^s({\bf x} - {\bf x}^\prime)= -1/[4\pi|{\bf x}-{\bf x}^\prime|]$, while ${\cal G}_{ij}^c({\bf x}-{\bf x}^\prime)= \partial_i^\prime\partial_jG^c({\bf x}-{\bf x}^\prime)$, where $G^c(({\bf x}-{\bf x}^\prime) = 1/(2\pi)\ln(|{\bf x}-{\bf x}^\prime|)$.  From these expressions, we find the nonzero components of the matrices $\hat{\cal G}^s$ and $\hat{\cal G}^c$ to be
\begin{equation}\label{eq:G_sphere_z}
{\cal G}^s_{xx}({\bf x}_n-{\bf x}_{n^\prime}) = {\cal G}^s_{yy}({\bf x}_n - {\bf x}_{n^\prime}) =
 -\frac{1}{4\pi}\frac{1}{|z_n-z_{n^\prime}|^3}
\end{equation} 
and 
\begin{equation}\label{eq:G_sphere_z2}
{\cal G}^s_{zz}({\bf x}_n-{\bf x}_{n^\prime}) = \frac{1}{2\pi}\frac{1}{|z_n-z_{n^\prime}|^3}
\end{equation}
for the case of a spheres in vacuum, and 
\begin{equation}\label{eq:G_rods_z}
{\cal G}^c_{yy}=-{\cal G}^c_{zz}=\frac{1}{2\pi}\frac{1}{|z_n-z_n^\prime|^2}.
\end{equation}
for the case of a cylinders in vacuum, with all other elements in both ${\cal G}^s$ or ${\cal G}^c$ equal to zero.

The matrix $\hat{t}$ is equal to 
\begin{equation}\label{eq:t_matrix}
\hat{t} = \hat{\delta\mu}(\hat{I}-\hat{\Gamma}\hat{\delta\mu})^{-1},
\end{equation}
where
\begin{equation}
\hat{\delta\mu} = \hat{\mu} - \hat{I},
\end{equation}
$\hat{I}$ is the identity matrix (since we assume that the host has permeability equal to unity) and the permeability tensor $\hat{\mu}$, discussed below, depends on the orientation of the magnetic field.

The demagnetization matrices $\hat{\Gamma}$ for spheres and cylinders in vacuum are also well known~\cite{Osborn1945}. For spheres,
\begin{equation}\label{eq:gamma_sphere}
\hat{\Gamma} = -\frac{1}{3}\hat{I}
\end{equation}
while for cylinders parallel to $\hat{z}$, the nonzero elements are
\begin{align}\label{eq:gamma_rods}
\Gamma_{yy} &= \Gamma_{zz} =-\frac{1}{2} \nonumber \\
\Gamma_{xx} &= 0.
\end{align}

If the applied magnetic field ${\bf H}_0$ and the saturation magnetization ${\bf M}_s$  are both parallel to $\hat{z}$, the permeability tensor $\hat{\mu}$ (in Gaussian units) takes the form~\cite{Pozar1990, Moorish1965}
\begin{equation}\label{eq:permeability_tensor}
\hat{ \mu} = \begin{pmatrix} \mu_1 & i \mu_2 & 0 \\
- i \mu_2 & \mu_1 & 0\\
0 & 0 & \mu_f 
\end{pmatrix}.
\end{equation}
If both ${\bf H}_0$ and ${\bf M}_s$ are parallel to  $\hat{x}$, then
\begin{equation}\label{eq:permeability_tensor_2}
\hat{ \mu} = \begin{pmatrix} \mu_f &  0&  0 \\
0 & \mu_1 & i\mu_2 \\
0 & -i\mu_2 & \mu_1 
\end{pmatrix}.
\end{equation}
In both cases, for YIG, the components of the permeability tensor $\hat{\mu}$ are given by
\begin{equation}\label{eq:mu1}
\mu_1 = \mu_f\left( 1+\frac{\omega_0 \omega_1}{\omega_0^2-\omega^2}\right)
\end{equation}
and 
\begin{equation}\label{eq:mu2}
\mu_2 = \mu_f \frac{\omega\omega_1}{\omega_0^2 - \omega^2},
\end{equation}
with $\omega_0 = \gamma | {\bf H}_0| - i\alpha\omega$ and $\omega_1 = 4\pi\gamma M_s/\mu_f$. Here $M_s$ is the magnitude of the saturation magnetization, $\gamma$ is the gyromagnetic ratio, and $\alpha$ is the phenomenological damping coefficient~\cite{Pozar1990}. For our purposes, we will assume~\cite{Cunha2015,Yu2014} that $4 \pi M_s = 1760\ G$ and $\gamma = 2.8 \ MHz/G$ and we take $\alpha = 5.0\times10^{-5}$. With these parameters and the equations given in Ref.~\citenum{Yu2014}, we calculate $\mu_f$ to be $1.40$. We use Gaussian units throughout this article.

The $\hat{t}$-matrices for the four cases presented here are easily found by combining the previous equations. For a chain of spheres along the $z$ axis with ${\bf H}_0 \| {\bf M}_s \| \hat{z}$ the non-vanishing components of the $\hat{t}$-matrix are found to be
\begin{align}\label{eq:t_matrix_sphere_z}
t_{xx}  &=  t_{yy}  =  \frac{\mu_1 -2 +\mu_1^2-\mu_2^2}{D} \nonumber \\
t_{xy}  &=  - t_{yx}  =  \frac{ 3 \mu_2}{D} \nonumber \\
t_{zz}  &= \frac{3(\mu_f -1)}{\mu_f +1} 
\end{align}
where $D  = \frac{1}{3}(4+4\mu_1+\mu_1^2-\mu_2^2)$.

For a chain of spheres along the $z$ with ${\bf H}_0 \| {\bf M}_s \| \hat{x}$ the non-vanishing components of the $\hat{t}$-matrix are
\begin{align}\label{eq:t_matrix_sphere_x}
t_{xx}&= \frac{3(\mu_f -1)}{\mu_f+1}\nonumber \\
t_{yy} &= t_{zz} =  \frac{\mu_1 -2 +\mu_1^2-\mu_2^2}{D} \nonumber \\
t_{yz}  &=  - t_{zy}  =  \frac{ 3  \mu_2}{D }
\end{align}
where $D$ is the same in Eq.~\eqref{eq:t_matrix_sphere_z}. 

For a chain of cylinders along $z$ with cylinder axes parallel to $\hat{x}$ (see Fig.~\ref{fig:one}) with ${\bf H}_0\|{\bf M}_s \| \hat{z}$ the non-vanishing components of the $\hat{t}$-matrix are
\begin{align}\label{eq:t_matrix_rod_z}
t_{xx} &=\frac{\mu_1^2-4\mu_1 -\mu_2^2 +3}{D_1} \nonumber \\
t_{yy} &= \frac{2( \mu_1-1)}{D_1} \nonumber \\
t_{zz} &= \frac{2(\mu_f-1)}{\mu_f+1} \nonumber \\
-t_{xy} &= t_{yx}= \frac{2  \mu_2}{D_1}
\end{align}
where $D_1= \mu_1 - 3$.  

Finally, for a chain of cylinders along $z$ with cylinder axes parallel to $\hat{x}$ with ${\bf H}_0 \| {\bf M}_s \| \hat{x}$ the non-vanishing components of the $\hat{t}$-matrix are
\begin{align}\label{eq:t_matrix_rod_x}
t_{xx} & = \mu_f -1 \nonumber \\
t_{yy} & =  \frac{-\mu_1^2 +\mu_2^2+2}{D_2} \nonumber \\
t_{zz} & =  \frac{\mu^2_1-4\mu_1 -\mu_2^2 +3}{D_2} \nonumber \\
t_{yz} &= -t_{zy}= \frac{2  \mu_2}{D_2}.
\end{align} 
where $D_2 = \frac{1}{2}(\mu_1^2-2 \mu_1-\mu_2^2-3)$.

Having calculated the elements of the $\hat{t}$-matrices, the $\hat{{\cal G}}$ matrices, and the demagnetization matrix $\hat{\Gamma}$, we can now calculate the dispersion relations using  Eq.~\eqref{eq:mag_moment3D} or~\eqref{eq:mag_moment2D}.

\begin{figure}[t]
\includegraphics[width=0.45\textwidth]{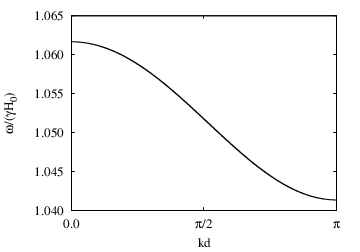} 
\caption{The real part of the calculated dispersion relations $\omega(k)$ of the transverse branch for a chain of YIG spheres parallel to $\hat{z}$, for the case ${\bf H}_0 \|{\bf M}_s \|\hat{z}$. We plot $\omega/(\gamma H_0)$ versus $kd$. The dispersion relation (full line) is obtained from Eq.~\eqref{eq:disp_sphere_z}, using the parameters $t_{xx}$ and $t_{xy}$ as given in Eq.~\eqref{eq:t_matrix_sphere_z}.  Only one real solution occurs for this geometry, as described in the appendix.  The solution is symmetric about k = 0.}
\label{fig:two}
\end{figure}  

\vspace{1.0in}
\section{Propagating Wave Solutions for Spherical and Cylindrical Particles}~\label{three}
\subsection*{ Spherical Particles}

Using Eqs.~\eqref{eq:mag_moment3D},~\eqref{eq:G_sphere_z},~\eqref{eq:G_sphere_z2}, and either~\eqref{eq:t_matrix_sphere_z} or~\eqref{eq:t_matrix_sphere_x},  we can obtain dispersion relations for propagating transverse ($T$) or longitudinal ($L$) waves for the cases ${\bf H}_0  \| {\bf M}_s \| \hat{z}$ and ${\bf H}_0 \| {\bf M}_s \| \hat{x}$.   In the former case, we find that the equations for the two transverse components of ${\bf m}$ are coupled, leading to solutions which are left- or right-circularly polarized waves.  The $L$ branch is
 found to be independent of $k$ in the quasistatic approximation.

To obtain the dispersion relations for the $T$ and $L$  branches, with ${\bf H}_0 \| {\bf M}_s \| \hat{z}$, we assume that ${\bf m}_n ={\bf m}_0 e^{ikn}$.   We can then write down the matrix equation for the two coupled $T$ branches as
\begin{equation}\label{eq:18}
{\bf m}_0 = \frac{2a^3}{3d^3}\begin{pmatrix} 
t_{xx} &  it_{xy}   \\
-it_{xy} & t_{xx} 
\end{pmatrix}{\bf m}_0Cl_3(kd).
\end{equation}
The equation for the $L$ branch is given as
\begin{equation}\label{eq:19}
{\bf m}_{0z} = -\frac{4 a^3}{3 d^3} t_{zz}{\bf m}_{0z}Cl_3(kd).
\end{equation}
In both these expressions,
\begin{equation}
Cl_s(z) = \sum^\infty_{n' =1,2,...}\frac{\cos{n' z}}{n'^s}
\nonumber
\end{equation}
is known as the Clausen function~\cite{Clausen1832}.  We have obtained these expressions by making a change of variables to rewrite the sum in Eq.~\eqref{eq:mag_moment3D} in terms of a single summand.  Since $t_{zz}$ is independent of $\omega$ we find that the $L$ branch is independent of wave number within the quasistatic approximation.

In principle the sum in the Clausen function can be evaluated numerically, which would allow the calculation of dispersion relations including the effects of all neighbors.  In practice this may be difficult, since the elements of both the $\hat{t}$-matrix and the Clausen function are multivalued functions of $\omega$ and of $kd$.  Hence, we will only include only the nearest-neighbor contributions for the remainder of this paper.

In the case of only nearest-neighbor interactions, i.\ e., including only the term $n' = 1$, the equation for the $T$ branches simplifies to
\begin{equation}\label{eq:disp_sphere_z}
\left[1- \frac{2a^3}{3d^3}\begin{pmatrix} 
t_{xx} &  it_{xy}   \\
-it_{xy} & t_{xx} 
\end{pmatrix}\cos{kd}\right]{\bf m}_0 = 0.
\end{equation}
The solutions to this equation, which are found by setting the determinant of the matrix of coefficients of ${\bf m}_0$ equal to zero, give the dispersion relations for the coupled transverse modes when the chain and magnetic field both lie in the $\hat{z}$ direction. Since the $\hat{t}$-matrix depends on the frequency $\omega$ this equation represents an implicit relation between $\omega$ and $k$ for the coupled waves.    In this geometry, as already mentioned, the two solutions are left- and right-circularly polarized $T$ waves. Since these $T$ waves have different dispersion relations, a linearly polarized $T$ wave will undergo a rotation as it propagates along the chain in a manner similar to the Faraday effect in a bulk homogeneous magnetic material~\cite{Fu2008}.

In Fig.~\ref{fig:two}, and all following figures, we plot the real part of the dispersion relation for the coupled waves for the various cases presented here.  In Fig.~\ref{fig:two} we plot the dispersion relations for the case ${\bf  H}_0 \| {\bf M}_s  \| \hat{z}$ (and also the chain parallel to $\hat{z}$) as functions of $\omega/(\gamma H_0)$, taking $H_0 = 1 \ T$ and $\gamma = 2.8$ MHz/Gauss.  As explained in the appendix, only a single solution exists for the geometry considered here and we have numerically verified that, upon changing the strength of the off-diagonal matrix element, we find two coupled solutions for the $T$ waves whose dispersion is dependent on the strength of the off-diagonal matrix element.  Therefore, the Faraday rotation angle per unit chain length, a measure of the difference in wave vector for circularly polarized waves, is also strongly dependent on the off-diagonal element.  The Faraday rotation angle is~\cite{Pike2016} $\theta(\omega) = \frac{1}{2}\left(k_1(\omega)-k_2(\omega)\right)$ where $k_1$ and $k_2$ are the wave numbers of the two circularly polarized solutions at frequency $\omega$.

Next, we consider the case where ${\bf H}_0 \| {\bf M}_s \| \hat{x}$, while the chain of magnetic spheres is again parallel to $\hat{z}$. In this case, the permeability tensor is given by Eq.~\eqref{eq:permeability_tensor_2} and the $\hat{t}$-matrix has components given by Eq.~\eqref{eq:t_matrix_sphere_x}. From the $\hat{t}$-matrix it is clear that the $T$ waves polarized in the $x$ direction are independent of $k$ within the quasistatic approximation, and are decoupled from waves polarized in the $y$ and $z$ directions. 

The coupled $y$ and $z$ components once again satisfy equations of motion which are analogous to Eq.~\eqref{eq:disp_sphere_z} but with the permeability tensor now given by Eq.~\eqref{eq:permeability_tensor_2}. As in the previous case  we can assume ${\bf m}_n ={\bf m}_0 e^{ikn}$ and, considering only nearest-neighbor coupling, can obtain the dispersion relations for the coupled $y$ and $z$ waves.  The matrix equation for these two components is found to be
\begin{equation}
\left[1 - \frac{2a^3}{3d^3} \cos{kd}\begin{pmatrix}
t_{yy} &  -2it_{yz}   \\
-it_{yz} & -2t_{yy} 
\end{pmatrix}\right] {\bf m}_0 = 0.
\end{equation}
Once again, we can determine the dispersion relations for the two coupled branches by setting the determinant of the matrix of coefficients of ${\bf m}_0$ equal to zero and solving for $k(\omega)$. 

\begin{figure}[t]
\includegraphics[width=0.45\textwidth]{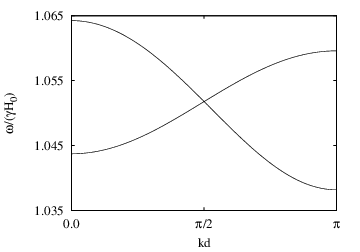} 
\caption{The real part of the calculated dispersion relations $\omega/(\gamma H_0)$ for a periodic array of spheres parallel to $\hat{z}$ with ${\bf H}_0 \| {\bf M}_s \| \hat{x}$. We showing the coupled $L$ and $T$ branches as functions of $kd$.  The third branch (not shown) is a pure $T$ mode whose frequency is independent of $k$ within the quasistatic approximation.  We use the same parameters as in Fig.~\ref{fig:two}.}
\label{fig:three}
\end{figure}

In Fig.~\ref{fig:three} we present the resulting dispersion relations $\omega(k)$ for a chain of YIG spheres oriented along the $\hat{z}$ axis with ${\bf H}_0\| {\bf M}_s \| \hat{x}$. In this geometry the two coupled $L$ and $T$ wave are elliptically polarized.  The frequency of the second $T$ branch (not shown) is independent of $k$ within the quasistatic approximation.

\subsection*{Cylindrical Particles}

Next, we consider the case of cylindrical particles whose long axes is parallel to $x$ in a chain which is periodically arranged parallel to $z$ (see Fig.~\ref{fig:one}).  Using Eqs.~\eqref{eq:mag_moment2D},~\eqref{eq:G_rods_z},~\eqref{eq:gamma_rods}, and either~\eqref{eq:t_matrix_rod_z} or~\eqref{eq:t_matrix_rod_x} we can determine the coupled equations for the dipole moments in either of two cases:  $ {\bf H}_0\|{\bf M}_s \| \hat{z}$ and ${\bf H}_0\| {\bf M}_s \| \hat{x}$.    In the former case, we find from the form of the $\hat{t}$-matrix that the frequency of the $L$  branch  (i.\ e., that polarized parallel to $\hat{z}$), is independent of $k$ while the two $T$ branches, which are parallel to the $x$ and $y$ axes, are coupled. The equation for these two coupled branches can be written as
\begin{equation}
{\bf m}_0 = \frac{a^2}{d^2}\begin{pmatrix} 
0 &  it_{xy}   \\
0 & -t_{yy} 
\end{pmatrix}{\bf m}_0 Cl_2(kd),
\end{equation}
or, if we include only nearest-neighbor interactions, i.e. $n' =1$,
\begin{equation}\label{eq:cyl_h0_z}
{\bf m}_0 = \frac{a^2}{d^2}\begin{pmatrix} 
0 &  it_{xy}   \\
0 & -t_{yy} 
\end{pmatrix}{\bf m}_0\cos{kd}.
\end{equation}

We may write out Eq.~\eqref{eq:cyl_h0_z} as two coupled algebraic equations:
\begin{equation}
\left[1+\frac{a^2}{d^2}t_{yy}(\omega)\cos(kd)\right]m_{y0} = 0
\label{eq:my0}
\end{equation}
and
\begin{equation}
m_{x0} - \frac{a^2}{d^2} it_{xy}(\omega)\cos(kd)m_{y0} = 0.
\label{eq:mxy0}
\end{equation}
This pair of equations may, in principle, have two solutions for each $k$, corresponding to the two possible transverse branches.  We consider each in turn.

For the first solution, $m_{y0} \neq 0$.  In that case, according to Eq.~\eqref{eq:my0}, the dispersion relation is determined by the implicit equation $1 + (a^2/d^2)t_{yy}\cos(kd) = 0$, since $t_{xx}$ (and $t_{xy}$) depend on $\omega$.   The $k$-dependent polarization of the corresponding mode is obtained from Eq.~\eqref{eq:mxy0} and satisfies $m_{x0}/m_{y0} = (a^2/d^2)it_{xy}\cos(kd)$.    Using $\cos(kd) = -(d^2/a^2)[1/t_{yy}]$, we obtain $m_{x0}/m_{y0} = -it_{xy}/t_{yy}$ along this branch.   Using Eq.~\eqref{eq:t_matrix_rod_z} for $t_{xx}$ and $t_{xy}$, we find that $m_{x0}$ and $m_{y0}$ are $\pi/2$ out of phase, so this wave is elliptically polarized.   

In the second solution, we take $m_{y0} = 0$, so the mode would be polarized along $\hat{x}$.   Eq.~\eqref{eq:my0} is then automatically satisfied, provided that $t_{yy}$ is finite.  In order for Eq.~\eqref{eq:mxy0} to be satisfied with $m_{y0} = 0$ and $m_{x0} \neq 0$, we must have $t_{xy} = \infty$ as well as $t_{yy}$ finite.    But if we consider the expressions given in Eq.~\eqref{eq:t_matrix_rod_z} for $t_{xy}$ and $t_{yy}$, we find that there exists no frequency for which both $t_{xy}(\omega) = \infty$ and $t_{yy}(\omega)$ is finite.  Therefore, we conclude that there is only a single propagating branch for this geometry, which is elliptically polarized, propagating along $\hat{z}$, and described by the dispersion relation $\omega(k)$ given implicitly by setting the quantity in square brackets in Eq.~\eqref{eq:my0} equal to zero.  

\begin{figure}[t]
\includegraphics[width=0.45\textwidth]{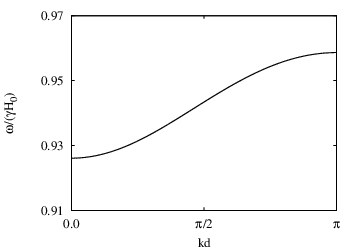} 
\caption{The real part of the calculated frequency, $\omega/(\gamma H_0)$, as a function of $kd$ for the elliptically polarized $T$ branch given by Eq.~\eqref{eq:my0} when the YIG cylinder axes is parallel to $\hat{x}$ and ${\bf H}_0 \|{\bf M}_s \| \hat{z}$. }
 \label{fig:four}
\end{figure}

In Fig.~\ref{fig:four} we show the dispersion relation for a chain of YIG cylinders oriented along the $x$ axis with ${\bf H}_0 \| {\bf M}_s \| \hat{z}$. In this geometry, and within the quasistatic approximation, as mentioned above, we find a branch elliptically polarized in the $xy$ plane.  There is also a $z$-polarized ($L$) mode, not shown in the figure, whose frequency is independent of $k$ and which is uncoupled to the $xy$ polarized branch. 

We now consider the final case, in which the long axis of the cylinders are aligned along the $x$ axis and ${\bf H}_0 \| {\bf M}_s \|\hat{x}$. In this case we notice from the $\hat{t}$-matrix that the $x$-polarized branch, which is one of the transverse branches, decouples from the $y$ and $z$ branches and has a frequency independent of $k$.  The remaining two branches obey, in the nearest-neighbor approximation, the equation
\begin{equation}
{\bf m}_0 = \frac{a^2}{d^2}\begin{pmatrix} 
-t_{yy} &  it_{yz}   \\
it_{yz} & t_{zz} 
\end{pmatrix}{\bf m}_0\cos{kd}.
\end{equation}
This equation can be solved by the methods described above to give the dispersion relations of the coupled $y$ and $z$ branches. 

\begin{figure}[t]
\includegraphics[width=0.45\textwidth]{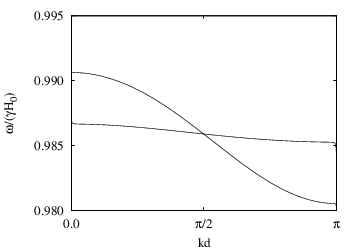} 
\caption{The real part of the calculated frequencies, $\omega/(\gamma H_0)$,  for the coupled $L$ and $T$ modes as functions of $kd$ when the axis of the YIG cylinders is parallel to $\hat{x}$  and ${\bf H}_0 \|{\bf M}_s \| \hat{x}$. }
\label{fig:five}
\end{figure}

In Fig.~\ref{fig:five} we show the dispersion relations for a chain of YIG cylinders oriented with their long axes parallel to the $\hat{x}$ axis with ${\bf H}_0 \|{\bf M}_s \| \hat{x}$.  In this geometry the coupled $L$ and $T$ waves are elliptically polarized.  The third branch, which is a $T$ branch, polarized parallel to $\hat{x}$,  is independent of wave number within the quasistatic approximation and is not plotted in the figure. 

\section{Discussion}~\label{four}

While we show numerical results only for nearest-neighbor interactions, we have found numerically that including further neighbors does not qualitatively change these results.  We do not show these numerical results in the paper, but have presented the relevant formal expressions in Eqs.~\eqref{eq:18} and~\eqref{eq:19}.   We also note that the spacing between the particles in our calculations was chosen to limit the effects of higher-order dipole moments.  If the particles are spaced closer together than about $a/d = 1/3$,  one must consider additional contributions from magnetic quadrupole and higher modes.  If these were included, the dispersion relations would quantitatively change, but the qualitative results would remain similar for the lowest bands.  We have also shown that there are usually several branches of magnon waves with dispersion relations that depend on the external magnetic field magnitude and orientation. These waves propagate along the chain with different polarization- and wave-number-dependent group velocities.

As mentioned earlier, these magnon waves will transmit power along the chains.   The energy density is proportional to the square of the absolute wave amplitude, and the transmitted power is therefore equal to the energy density multiplied by the group velocity of the wave.   The group velocity $v_{gn}(k)$ of the n$^{th}$ polarization can be computed from the dispersion relation via the relation $v_{ng}(k) = d\omega_n(k)/dk$.    Just as for plasmonic waves on metallic particle chains, the transmitted power can be controlled by introducing geometries such as T-junctions, which allow the power to be split into two parts with relative magnitudes depending on frequency~\cite{Brongersma2000}.   The calculations of the splitting will be more complicated, however, because the waves may be circularly or elliptically polarized.  Similarly, one can calculate the magnetization current due to these waves.

The dispersion relations of magnons on chains of either spheres or cylinders calculated here include the effects of Gilbert damping to describe the relaxation of the magnetic moments within each particle. Additionally, one could also include the effects of a magnetic torque, which would further couple the three polarization modes, or one could include the effects of crystalline anisotropy or demagnetization fields.  Since both crystalline anisotropy and demagnetization fields are frequency independent they would directly affect the permeability matrix elements  $\mu_1$ and $\mu_2$~\cite{Moorish1965}.  Therefore, the effects of these terms will quantitatively change the dispersion relations calculated here, but will not affect the qualitative results.  In addition, since the crystal anisotropy is a sensitive function of temperature~\cite{Moorish1965}  one can modify the dispersion relations for different polarizations in a controllable manner by varying the temperature.   It should also be possible to include effects beyond the quasistatic approximation by extending the approach of Weber and Ford~\cite{Weber2004} to magnon waves propagating via magnetic dipole-dipole interactions.

To summarize, we have calculated the dispersion relations for magnon waves propagating along chains of YIG spheres and cylinders, in which only the quasistatic coupling between magnetic dipoles is included.   We found that, depending on the orientation of the static magnetization and applied magnetic field relative to the chain, these waves are either circularly, elliptically, or linearly polarized.   In the first two cases, an incident linearly polarized wave will undergo Faraday rotation, analogous to that seen in bulk magnetic compounds~\cite{Fu2008}, and this rotation can be tuned in a controllable way.   These waves also carry magnetization current (magnon current) along the chain.  Thus, it should be possible to use chains of YIG particle to transmit power wirelelssly in a meso- or nanoscale circuit without generation of large electric potentials and within a diameter small compared to the wavelength,

\begin{acknowledgments}
N. P was supported by the Belgian Fonds National de la Recherche Scientifique FNRS under grant number PDR T.1077.15-1/7 and both authors acknowledge funds from the Center for Emerging Materials at The Ohio State University, an NSF MRSEC (Grant No.\ DMR-1420451).
\end{acknowledgments}

\appendix*
\section{Transverse Modes on a Chain of YIG Spheres}
For the case of a chain of spheres along the $z$ axis with ${\bf H}_0 \| {\bf M}_s \| \hat{z}$ (see Section~\ref{three})  we found that the coupled $T$ modes propagate if the determinant of the matrix of coefficients of ${\bf m}_0$ [Eq.~\eqref{eq:disp_sphere_z}] vanishes. Evaluating this determinant and setting it equal to zero gives the following solutions for the two $T$ modes in the nearest-neighbor approximation:
\begin{equation}\label{eq:sphere_z_solution}
\frac{2 a^3}{3d^3}\cos{kd} = \frac{1}{t_{xx}\pm  t_{xy}}. 
\end{equation}

We can use Eq.~\eqref{eq:t_matrix_sphere_z} for the values of $t_{xx}$ and $t_{xy}$ to determine the dependence of the coupled modes on the permeability matrix elements as
\begin{equation}\label{eq:appendix_a}
\frac{1}{t_{xx}\pm  t_{xy}}  = \frac{1}{3}+ \frac{1}{\mu_1\pm \mu_2 -1}.
\end{equation}
Furthermore, we can use Eqs.~\eqref{eq:mu1} and~\eqref{eq:mu2} to further simplify the equations and explain why, for fixed magnetic field and wave vector
${\bf k} = k\hat{z}$, there is only one propagating $T$ mode.

Substituting in Eqs.~\eqref{eq:mu1} and~\eqref{eq:mu2} into the quantity $\mu_1 \pm \mu_2$ we find that 
\begin{equation}\label{eq:appendix3}
\mu_1 \pm \mu_2 =\mu_f \left(1-\frac{\omega_1}{\omega \pm \omega_0}\right)
\end{equation}
and therefore, by combining Eqs.~\eqref{eq:sphere_z_solution},~\eqref{eq:appendix_a}, and~\eqref{eq:appendix3}  we find that the dispersion relationship becomes
\begin{equation}\label{eq:appendix4}
\frac{2 a^3}{3d^3}\cos{kd} =  \frac{1}{3}+ \frac{1}{\mu_f-1 +\frac{\mu_f \omega_1}{\omega\pm\omega_0}}.
\end{equation}

Eq.~\eqref{eq:appendix4} presents an analytic solution for the $T$ waves.  In general, one obtains a solution for this equation by solving for $\omega(k)$ [or $k(\omega)$]. If we neglect damping (by taking $\omega_0$ to be real), then for a given value of $k$, the $+$ and $-$ signs in Eq.~\eqref{eq:appendix4} give two values of $\omega$, which are equal and opposite (i.\ e., one is positive and one is negative).   These solutions correspond to waves propagating with the same frequency but with wave vectors $k\hat{z}$ and $-k\hat{z}$.   Thus, for a given value of the one-dimensional vector ${\bf k}$, there is actually only one solution.   We can also obtain the corresponding form of the vector ${\bf m}_0$ from Eq.~\eqref{eq:disp_sphere_z}.   It is readily found that the $\pm$ solutions correspond to $m_{0y} = \mp im_{0x}$. These solutions represent circularly polarized waves propagating in the +$\hat{z}$ and $-\hat{z}$ directions, respectively.  Thus, in summary, for each positive or negative value of $k$, there is one one propagating, circularly polarized wave with frequency $\omega(k)$. If we allow $\omega_0$ to be complex, as is actually the case, then we find that the $+k$ and $-k$ solution are each damped as they travel along $+\hat{z}$ and $-\hat{z}$. In Fig.~\ref{fig:two} we have plotted the real part of the complex dispersion relation given by Eq.~\eqref{eq:disp_sphere_z}, using the parameters given below Eq.~\eqref{eq:mu2}.

\end{document}